\documentclass[10pt]{iopart}

\usepackage{iopams}
\usepackage{graphicx}
\usepackage{epstopdf}

\begin{document}

\title{Multi-resolution Progressive Computational Ghost Imaging}
\author{Cheng Zhou$^{1,2,3}$, Tian Tian$^4$, Chao Gao$^5$, Wenli Gong$^{6}$, and Lijun Song$^{1,2}$}

\address{$^1$ Institute for Interdisciplinary Quantum Information Technology, Jilin Engineering Normal University, Changchun 130052, China\\
$^2$ Jilin Engineering Laboratory for Quantum Information Technology, Changchun 130052, China\\
$^3$ Center for Quantum Sciences and School of Physics, Northeast Normal University, Changchun 130024, China\\
$^4$ School of Science, Changchun University, Changchun 130022, China\\
$^5$ Department of Physics, Changchun University of Science and Technology, Changchun 130022, China\\
$^6$ Key Laboratory for Quantum Optics and Center for Cold Atom Physics of CAS, Shanghai Institute of Optics and Fine Mechanics, Chinese Academy of Sciences, Shanghai 201800, China}
\ead{gongwl@siom.ac.cn and ccdxslj@126.com}

\begin{abstract}
Ghost imaging needs massive measurements to obtain an image with good visibility and the imaging speed is usually very low. In order to realize real-time high-resolution ghost imaging of a target which is located in a scenario with a large field of view (FOV), we propose a high-speed multi-resolution progressive computational ghost imaging approach. The target area is firstly locked by a low-resolution image with a small number of measurements, then high-resolution imaging of the target can be obtained by only modulating the light fields corresponding to the target area. The experiments verify the feasibility of the approach. The influence of detection signal-to-noise ratio on the quality of multi-resolution progressive computational ghost imaging is also investigated experimentally. This approach may be applied to some practical application scenarios such as ground-to-air or air-to-air imaging with a large FOV.
\end{abstract}

\vspace{2pc}

\section{Introduction}

Ghost imaging (GI) is realized by correlating the light field reflected (or transmitted) from the object with the reference light field \cite{1995s}. Because all the photons transmitted (or reflected) from the object illuminate the same bucket detector, this technique has the superiority of high sensitivity in detection and high efficiency in information extraction \cite{hshegi}, and in recent years, it has aroused increasing interest in the applications like remote sensing \cite{rese1,rese2}, biomedical imaging \cite{bigc,x-ray1,x-ray2,x-ray3}, super-resolution \cite{suregi}, optical encryption \cite{open1}. However, GI needs massive measurements to obtain an image with good visibility, which is difficult to realize real-time imaging \cite{realtime}. Although the measurements can be decreased by some image reconstruction algorithms such as compressive ghost imaging, the optimization procedure of image restoration is time-consuming and the computation resource required is relatively high, especially for a scene with a large field of view (FOV) and high resolution \cite{fovhr}, which also hindered its application. In some practical application scenarios like ground-to-air or air-to-air imaging with a large FOV, the target only occupies a small region in the whole FOV and high-resolution imaging is required only for the target area, for which the imaging speed usually should be up to 25Hz. Therefore, in order to realize real-time imaging, the measurements should be obviously decreased and the image reconstruction had better be a quick linear algorithm.

Recently, some solutions have been proposed. Zhou et. al\cite{hcgi1} distinguished the high and low resolution regions in the large FOV by low resolution speckle field and utilized the hybrid-scale speckle pattern to obtain the high quality image. On the down side, the method defaults to high resolution outside the low resolution recognition area. As we all know, for imaging of sparse scene in the large FOV, high-speed and high-resolution causes huge resource consumption in data acquiring, storage, delivery and processing. Hence, the work neglected the resources waste of high resolution imaging caused by the sparse target in the large FOV. Sun et. al\cite{mcgi} realized the adaptive ghost imaging in large FOV based on multi-scale speckle. The method first finds out the location of interested objects before obtaining all the data, then achieves image of those parts with high resolution, which is more effective than Ref.\cite{hcgi1}. However the work shows a great dependence on the edge recognition algorithm and need repeated measurements and recalculation after resolution adjustment. For this reason, a more efficient multi-resolution imaging without additional detection is needed.

If the target area can be locked by a low-resolution image with a small number of measurements, then we only need to modulate the light fields corresponding to the target area, and the imaging speed can be obviously improved when an optimized coded illumination source is used. So in this paper, we propose a multi-resolution progressive computational ghost imaging (MPGI) method which is rooted in the application of Hadamard transform\cite{oh} and the ideas of progressive transmission\cite{pr2000}.The reordered Hadamard derived pattern is utilized in processings of low resolution location and multi-resolution imaging. Since Hadamard matrix is orthogonal and symmetric, the effect of the correlated noise can be eliminated, and the image of the target can be reconstructed accurately. Meanwhile, the progressive transmission plays an important role that it's widely used in image transmission and geographical information system. And it allows one to transmit the indispensable data without any redundancy to enlarge the response speed. Based on this idea, the MPGI method can get multi-resolution images without additional detection. The correctness and feasibility of the method is verified by experiments. In order to provide a clear guidance of practical application, we also discuss the influence of detection signal-to-noise ratio on imaging through practical experiments. The results show that this method can be applied to high-speed and high-resolution imaging with the large FOV.

\section{ Image reconstruction and multi-resolution progressive computational ghost imaging}
\label{HTGI}

\begin{figure}[htbp]
\centering
\includegraphics[width=8cm]{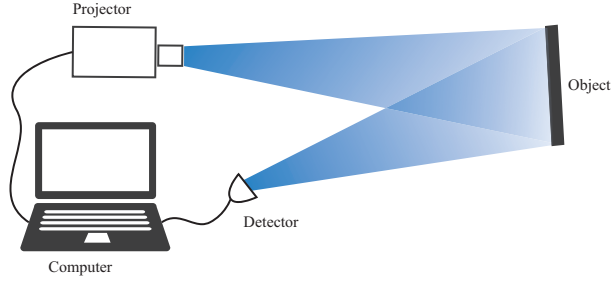}
\caption{(Color online) Schematic diagram of computational ghost imaging.}
\label{scheme}
\end{figure}

The schematic diagram of computational ghost imaging (CGI) is presented in Fig.~\ref{scheme}. A projector as the light modulator projects the modulated light field onto an object with a reflection coefficient $O(x,y)$. The total reflected signals are collected by a bucket detector. The $m$th light field and bucket signal are expressed as $I^{(m)}(x,y)$, $B^{(m)}$, respectively.

\subsection{Image reconstruction of ghost imaging}

In conventional GI, the reflected coefficient can be obtained by computing the correlation between $I^{(m)}(x,y)$ and $B^{(m)}$

\begin{eqnarray}\label{eqtgi}
O_{\textrm{GI}}(x,y) &= \frac{1}{M}\sum^{M}_{m=1}[(B^{(m)}-\langle B^{(m)}\rangle)\cdot I^{(m)}(x,y)]\\
&= \frac{1}{M}\sum^{M}_{m=1}[(B^{(m)}-\langle B^{(m)}\rangle)(I^{(m)}(x,y)-\langle I^{(m)}(x,y)\rangle)],
\end{eqnarray}
where $\langle B^{(m)}\rangle=\frac{1}{M}\sum^{M}_{m=1}B^{(m)}$ and
$\langle I^{(m)}(x,y)\rangle=\frac{1}{M}\sum^{M}_{m=1}I^{(m)}(x,y)$. We can transform the $I^{(m)}(x,y)$ (dimensions $p\times q$) of M measurements into a matrix form

\begin{equation}\label{PHI}
\Phi=\left[
       \begin{array}{cccc}
         I^{(1)}(1,1) & I^{(1)}(1,2) & \cdots & I^{(1)}(p,q) \\
         I^{(2)}(1,1) & I^{(2)}(1,2) & \cdots & I^{(2)}(p,q)  \\
         \vdots & \vdots & \ddots & \vdots  \\
         I^{(M)}(1,1) & I^{(M)}(1,2) & \cdots & I^{(M)}(p,q) \\
       \end{array}
     \right].
\end{equation}
here, each row of the matrix $\Phi$ is converted from a row vector of length $p\times q$, which is obtained by reshaping the $m$th light field $I^{(m)}(x,y)$.

Thus, Eq.~\ref{eqtgi} can be rewritten into a matrix form

\begin{eqnarray}\label{mgi}
O_{\textrm{GI}}(x,y) &= \frac{1}{M}(\mathbf{\Phi}-\textbf{\textit{I}}\langle \mathbf{\Phi} \rangle)^{T}(\textbf{\textit{B}}-\textbf{\textit{I}}\langle \textbf{\textit{B}}\rangle)\\
&= \frac{1}{M}(\mathbf{\Phi}-\textbf{\textit{I}}\langle \mathbf{\Phi}\rangle)^{T}(\mathbf{\Phi}-\textbf{\textit{I}}\langle \mathbf{\Phi}\rangle)\textbf{\textit{O}}\\
&= \frac{1}{M}\mathbf{\Psi}^{T}\mathbf{\Psi}\textbf{\textit{O}},
\end{eqnarray}
where, $\textit{\textbf{B}}$ is an $M\times 1$ vector and composed by M times of measurement, that is $\textit{\textbf{B}}=[B^{(1)},B^{(2)},\cdots,B^{(M)}]$. Similarly, $\textbf{\textit{O}}$ is an $M\times 1$ vector made up of the reflection coefficient $O(x,y)$ of object, $\textbf{\textit{O}}=[O(1,1),O(1,2),\cdots,O(p,q)]^T$. In addition, $\mathbf{\Psi}=\mathbf{\Phi}-\textbf{\textit{I}}\langle\mathbf{\Phi}\rangle$, $\mathbf{\Phi} \textit{\textbf{O}}=\textit{\textbf{B}}$, and $\langle \textit{\textbf{B}}\rangle=\langle\mathbf{\Phi}\rangle\textit{\textbf{O}}$. $\textit{\textbf{I}}$ represents an $M\times1$ column vector of all elements with a value of 1. $\langle\mathbf{\Phi}\rangle$ is a $1\times N$ row vector, which denotes the average of each column of $\mathbf{\Phi}$. In theory, a high quality reconstructed image will be obtained by Eq.~\ref{mgi} if $\mathbf{\Psi}^{T}\mathbf{\Psi}$ is a diagonal matrix. And in CGI system, the $\mathbf{\Psi}^{T}\mathbf{\Psi}$ full-width of diagonal nonzero elements at half-maximum is equal to the resolution of the preset light field, which is proportional to the resolution on the object plane. Hence, to realize the multi-resolution progressive computational ghost imaging, a multi-resolution $\mathbf{\Psi}^{T}\mathbf{\Psi}$ with different measurements number is extremely essential.

\subsection{Multi-resolution progressive computational ghost imaging}
\label{prcgi}
To enable high-speed multi-resolution progressive computational ghost imaging, we optimized the sequence of Hadamard derivative pattern to achieve real-time high-resolution imaging. In this section, we will show how to select and use Hadamard derived pattern, as it actualize the multi-resolution progressive image.

The Hadamard basis is a square matrix composed of $+1$ and $-1$, and can be generated rapidly by Kronecker product, that is

\begin{equation}
\left\{
   \begin{array}{l}
   H_{2^1}=\left[\begin{array}{cc}
                  +1 & +1 \\
                  +1 & -1
           \end{array}\right],\\
   H_{2^k}=H_{2^1}\otimes H_{2^{k-1}}=\left[\begin{array}{cc}
                                               +H_{2^{k-1}} & +H_{2^{k-1}} \\
                                               +H_{2^{k-1}} & -H_{2^{k-1}}
                                      \end{array}\right],\\
   \end{array}
   \right.
\end{equation}
for $2<k$ (integer),  where $ \otimes $ denotes the Kronecker product.
Hence, a Hadamard matrix of size $ M\times N$($M=N$),
\begin{equation}
H_{2^k}(m,n)=\left[\begin{array}{cccc}
               H{(1,1)} & H{(1,2)} & \cdots & H{(1,N)} \\
               H{(2,1)} & H{(2,2)} & \cdots & H{(2,N)} \\
                \vdots& \vdots & \ddots & \vdots \\
               H{(M,1)} & H{(M,2)} & \cdots & H{(M,N)}
             \end{array}\right].
\end{equation}

To construct a modulation matrix of the light field for GI, we will select an arbitrary row of $H_{2^k}(m,n)$ to obtain a two-dimensional Hadamard derived pattern $H_{2^k}^{(m)}(x,y)$.

\begin{figure}[htbp]
\centering
\includegraphics[width=8cm]{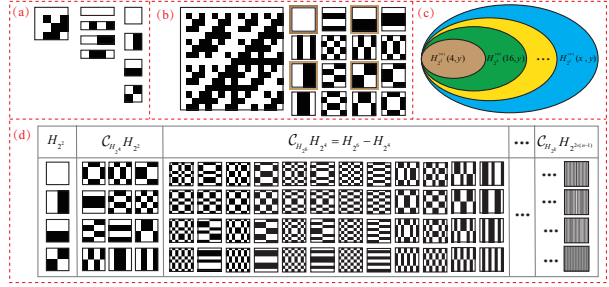}
\caption{(Color online) Reordering methods for Hadamard derived pattern. (a) Hadamard derived pattern of order four construction process; (b) Hadamard pattern of order eight and the corresponding derived pattern; (c) Hadamard derived pattern of different order sets; (d) Reordered Hadamard derived pattern.}
\label{hdpr}
\end{figure}

To get a set of Hadamard derived pattern $H_{2^k}^{(m)}(x,y)$, we first generate a high-order Hadamard matrix $H_{2^k}(m,n)$ that can satisfy the detection accuracy[taking $k=2$ for example, as shown in the left panel of  Fig.~\ref{hdpr}(a)]. Then, we extract each row of $H_{2^k}(m,n)$ to obtain the corresponding single row vectors[as shown in the middle panel of Fig.~\ref{hdpr}(a)]. At last, we acquire the two-dimensional matrix $H_{2^k}^{(m)}(x,y)$ of $x$ rows and $y$ columns. From up to down of the right panel of Fig.~\ref{hdpr}(a) we give the derived pattern $H_{2^2}^{(m)}(2,2)$ with $m=1,2,3,4$, respectively.

Since the Hadamard matrix is a direct product of $H_{2^1}$, the higher order Hadamard matrix naturally contains the distribution information of lower order Hadamard matrix. For example, with $k=4$, the 16pixel$\times$16pixel Hadamard pattern $H_{2^4}(16,16)$ is shown in the left subfigure of Fig.~\ref{hdpr}(b). The corresponding Hadamard derived pattern in the right subfigure Fig.~\ref{hdpr}(b) contains the 4-order Hadamard derived matrix $H_{2^2}^{(m)}(4,4)$, which is labelled by brown frames. For the purpose of reordering, we enlarge the latter matrix to the same size as the former one. In a similar derived approach, every pattern $H_{2^k}^{(m)}(x,y)$ is a proper subset of the $H_{2^{k'}}^{(m)}(x,y)$ for $k'>k$, and the relationship between the sets is shown in Fig.~\ref{hdpr}(c).

With the lower and higher derived pattern in our hand, we are now ready to reorder them. For a high order derived pattern, we first extract the lowest order derived pattern, i.e., $H_{2^2}^{(m)}(2,2)$ and prepose it, and deal with the rest lower order pattern according to priority in the same way.

\begin{figure}[htbp]
\centering
\includegraphics[width=8cm]{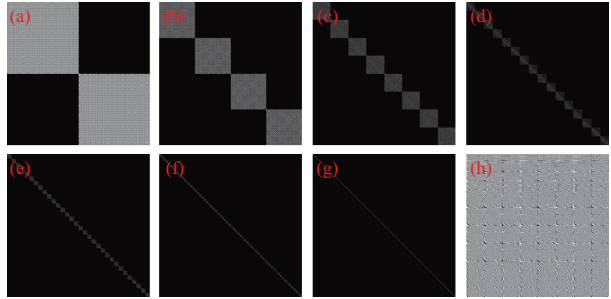}
\caption{ (Color online) Comparison of the full-widths at half-maximum of $\mathbf{\Psi}^{T}\mathbf{\Psi}$ with different measurements number. (a) $M=2^{2\times 1}$ ($2\times 2$ resolution); (b) $M=2^{2\times 2}$ ($4\times 4$ resolution); (c) $M=2^{2\times 3}$ ($8\times 8$ resolution); (d) $M=2^{2\times 4}$ ($16\times 16$ resolution); (e) $M=2^{2\times 5}$ ($32\times 32$ resolution); (f) $M=2^{2\times 6}$ ($64\times 64$ resolution); (g) $M=2^{2\times 7}$ ($128\times 128$ resolution); (h) Reordered Hadamard pattern.}
\label{diag}
\end{figure}

Concretely, in the language of the sets theory, $H_{2^2}$ is a proper subset of $H_{2^4}$ which is notated as $H_{2^2}\subsetneqq H_{2^4}$ [to make it simple, we omit the superscript $m$ in what follows], and the complementary set as $\complement_{H_{2^4}}H_{2^2}=H_{2^4}-H_{2^2}$, which is show by the green area in Fig.~\ref{hdpr}(c). As for the higher order, every two adjacent even orders will produce a corresponding complementary set $\complement_{H_{2^{2\times \kappa}}}H_{2^{2\times (\kappa-1)}}=H_{2^{2\times \kappa}}-H_{2^{2\times (\kappa-1)}}$, whose elements will be reordered in our scheme[as shown in Fig.~\ref{hdpr}(d)]. In a word, by enlarging the complementary set constantly we completed the Hadamard derived pattern reordering, $H_{2^{k}}^{(m)}(x,y)_{\textrm{new}}$, and the two dimensional form can be expressed as $\mathbf{H_{2^{k}}(m,n)_{\textrm{new}}}$ which is equivalent to $\mathbf{\Phi}$ in Eq.~\ref{PHI}. At this point, $\mathbf{\Psi}= \mathbf{H_{2^{k}}(m,n)}_{\textrm{new}}-\textbf{\textit{I}}\langle \mathbf{H_{2^{k}}(m,n)_{\textrm{new}}}\rangle$, the different full-widths at half-maximum of $\mathbf{\Psi}^{T}\mathbf{\Psi}$ with different measurements number $M=2^{2\times \kappa}(\kappa=1,2,\cdots)$ can be efficiently calculated, as shown in Fig~\ref{diag}. Relying on this, we can achieve the low-resolution location and multi-resolution progressive imaging quickly.

\section{Experimental results}
\label{er}
In ground-to-air or air-to-air situation a target object usually occupies a relatively small region in a large FOV. To enhance the efficiency, MPGI locates the small target object initially, and then gets the multi-resolution images at high speed.

\begin{figure}[htbp]
\centering
\includegraphics[width=8cm]{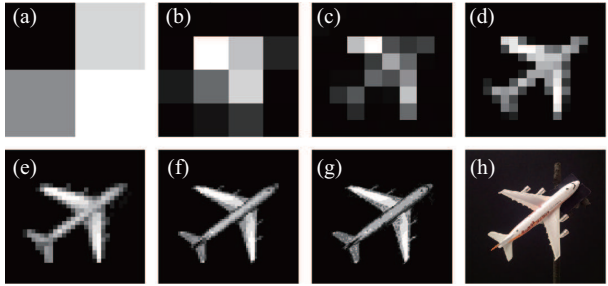}
\caption{Multi-resolution progressive computational ghost imaging results. (a) $M=2^{2\times 1}$ ($2\times 2$ resolution); (b) $M=2^{2\times 2}$ ($4\times 4$ resolution); (c) $M=2^{2\times 3}$ ($8\times 8$ resolution); (d) $M=2^{2\times 4}$ ($16\times 16$ resolution); (e) $M=2^{2\times 5}$ ($32\times 32$ resolution); (f) $M=2^{2\times 6}$ ($64\times 64$ resolution); (g) $M=2^{2\times 7}$ ($128\times 128$ resolution); (h) Original object.}
\label{pcgi}
\end{figure}

To verify the feasibility of the approach, we conducted a series of experiments. In the experiments, the object is an aircraft model [see Fig.~\ref{pcgi}(h)] with the size of $20cm\times 17cm$ and positioned about 0.72m, 2.42m away from the projector (XGIMI Z4 Air miniature projector) and bucket detector (Thorlabs, PDA100A-EC, 320-1100 nm, 2.4 MHz BW, 100 mm$^2$), respectively.

We adopted a set of reordered Hadamard derived pattern $H_{2^{2\times 7}}^{(m)}(128,128)_{\textrm{new}}$, $m=1,2,3,\cdots,16384$, i.e., we have done a set of seven resolutions MPGI experiments and the results are shown in Fig.~\ref{pcgi}. With the increase of the number of the measurements, the aircraft image information of the reconstructed image gradually becomes clear, i.e., image resolution steadily increased with measurement times. When the imaging resolution of MPGI is $4\times 4$ [Fig.~\ref{pcgi} (b) $M=2^{2\times 2}$], the position of the target object can be clearly identified. Hence, the target location can be locked by a low-resolution image with a small number of measurements. Even for $M=2^{2\times 5}$ ($32\times 32$ resolution), we can clearly distinguish the clear outline of the target object (an aircraft), which is enough for the military reconnaissance. As the number of measurements is further increased, the details of the reconstructed image gradually emerge. For example, the engines on both sides of the aircraft have been reconstructed by measuring $2^{2\times 6}$ times, as shown in Fig.~\ref{pcgi}(f). To achieve the same efficiency, 5460 times are needed in the conventional schemes. Furthermore, after a 4-fold increase in resolution, we find out that the reconstructed image Fig.~\ref{pcgi}(g) is much the same with Fig.~\ref{pcgi}(f), which shows that, in a few cases, high imaging resolution is indispensable to CGI, but there is a waste of resources in ultra-high imaging resolution. The results of MPGI experiments verify the feasibility of the multi-resolution progressive imaging and low resolution location.

\begin{figure}[htbp]
\centering
\includegraphics[width=8cm]{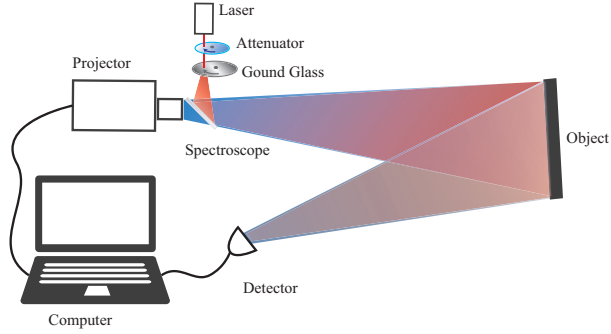}
\caption{(Color online) Schematic diagram of computational ghost imaging under background light noise.}
\label{fig1noi}
\end{figure}

To evaluate the performance of our scheme under background light noise, we introduce the detection signal-to-noise ratio (DSNR) which is defined as
\begin{equation}
DSNR=10log_{10}\frac{\langle B\rangle}{\sqrt{\langle(E-\langle E\rangle)^2\rangle}},
\end{equation}
where $\langle B\rangle$ is the mean signal power and $\langle E\rangle$ is the mean background light noise power\cite{prdsnr}.
For the 16384 measurements [Fig.~\ref{pcgi}], the DSNR is close to positive infinity (without noise), which exceeds the criterion in applicable large FOV. Therefore, we add a laser and modulate it by a rotating ground glass (as shown in Fig.~\ref{fig1noi}), and we weaken the intensity of the laser by attenuator to get different DSNR so as to discuss the multi-resolution progressive imaging performance of the proposed scheme.

\begin{figure}[htbp]
\centering
\includegraphics[width=8cm]{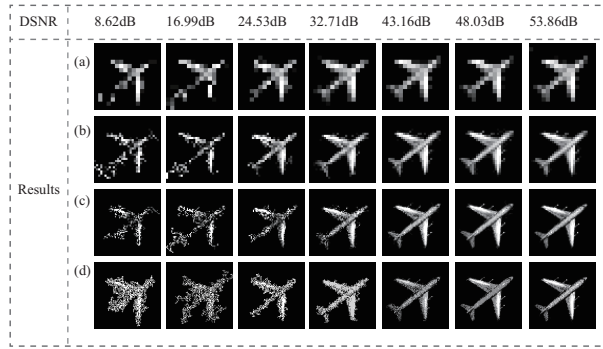}
\caption{Results of different DSNR. (a) $M=2^{2\times 4}$ ($16\times 16$ resolution); (b) $M=2^{2\times 5}$ ($32\times 32$ resolution); (c) $M=2^{2\times 6}$ ($64\times 64$ resolution); (d) $M=2^{2\times 7}$ ($128\times 128$ resolution).}
\label{rdsnr}
\end{figure}

In Fig.~\ref{rdsnr}, we show our MPGI results for different DSNR and resolution (equal to the measurement times). In the low DSNR case (DSNR<32.71dB), the low-resolution images can be effectively reconstructed [Fig.~\ref{rdsnr}(a)-(b)]. Unlike low resolution imaging, the high-resolution imaging can only get blurred reconstructed images [Fig.~\ref{rdsnr}(c)-(d)]. And if the DSNR is too low (DSNR<8.62dB), the experimental system will be invalid. By contrast, when the DSNR is high enough (DSNR>32.71dB), the method  can effectively reconstruct the multi-resolution images [Fig.~\ref{rdsnr}(a)-(d)], and it has fine image formation ability. And when the DSNR is 53.86dB, the image quality with M = 16384 [Fig.~\ref{rdsnr}(d) $128\times 128$ resolution] and M=4096 [Fig.~\ref{rdsnr}(c) $64\times 64$ resolution] are almost equal in vision. The results show that high-resolution images are difficult to obtain at low DSNR (i.e., the greater the resolution, the worse the noise resistance). When the DSNR is sufficient and there is no big demand for high-resolution, a large number of measurement times are not required to obtain the higher resolution images. Moreover, a high resolution imaging is still the optimal choice in the case of high DSNR. Obviously, the analysis of DSNR and multi-resolution imaging have a certain reference value for practical application.

\section{Discussion}
\label{Dis}
We propose and experimentally demonstrate a multi-resolution progressive computational ghost imaging method. First of all, we can use the echo signal and low resolution imaging to get the target area, the target area can be easily locked through low resolution images. Therefore, the MPGI method can be used for fast target location in sparse scenes. And then, we modulate the reordered Hadamard derived pattern light fields corresponding to the target area. Obviously, this can improve the imaging speed with a large FOV. Secondly, in the process of low resolution positioning and high resolution imaging, We get multiple multi-resolution images by using the Hadamard derived pattern of optimized coded  illumination source. If a ultra high speed spatial light modulator (such as digital micromirror devices or LED-array\cite{ledarray}) is used, the MPGI method can approximate the real-time imaging requirements of video frame rate 25Hz. Thirdly, comparing with the available technologies, MPGI method makes the GI technology more flexible. For example, 1. We overcame the difficulty of determining the spatial resolution; 2. We realized a on demand image resolution determined controllable image time.

However, for MPGI method, some problems still exist: the resolution can only increase exponentially. On the other hand, the anti-noise capability is poor. In the future, we can extend the transverse and vertical direction multi-resolution progressive imaging ability by deeply optimizing the Hadamard sequence, and combine some algorithms and techniques to improve the anti-noise ability and image quality.

\section{Conclusion}
\label{con}

In this paper, we discuss how to realize a multi-resolution progressive computational ghost imaging. We apply the Hadamard derived pattern to complete the high-speed multi-resolution progressive imaging, and provide a way for real-time ghost imaging with a large FOV. Due to the orthogonality of different rows in Hadamard derived pattern, the negative effects of noise is dramatically suppressed. Moreover, we also perform the experiment to accomplish MPGI of an aircraft model and investigate the effect of DSNR for different resolution which is equal to the measurement times in our work. We find that a higher resolution is combined with a high DSNR for an ideal image reconstruction. However, a lower DSNR does not always need a high resolution. This method may be applied to some
practical application scenarios with sparse targets.

\section*{Funding}
This work is supported by the Project of the Science and Technology Department of Jilin Province (Grant No. 20170204023GX); the Special Funds for Provincial Industrial Innovation in Jilin Province (Grant No. 2018C040-4); Youth Innovation Promotion Association of the Chinese Academy of Sciences; the Young Foundation of Science and Technology Department of Jilin Province (Grant No. 20170520109JH); and the Science Foundation of Education Department of Jilin Province (Grant No. 2016286).

\section*{Acknowledgments}
We thank Z. H. Wang, A. J. Sang and X. F. An for their fruitful discussions.

\section*{References}

  

\end{document}